# Complexity Analysis of Wind Energy, Wind Speed and Wind Direction in the light of nonlinear technique

**Sayantan Chakraborty[a*], Sourav Samanta[b], Shukla Samanta[c], Dipak Ghosh[c], Kumardeb Banerjee[b]**

[a] Department of Electrical Engineering, Dr. Sudhir Chandra Sur Institute of Technology & Sports Complex, JIS Group of Educational Initiatives, Kolkata – 700074, India

[b] Department of Instrumentation and Electronics Engineering, Jadavpur University, Kolkata – 700098, India

[c] Sir C.V. Raman Centre for Physics & Music, Jadavpur University, Kolkata – 700032, India

*sayantan.a2z@gmail.com

**Abstract:**

Wind energy has an inherent intermittent character due to certain inevitable factors of nature, such as availability of wind at different weather conditions, wind direction etc. To study the intermittent character of wind energy, its daily data along with the two other important quantities, wind speed and wind direction measured in a "showcase" wind farm for a span of ten years are analyzed applying a nonlinear robust tool Multifractal Detrended Cross-correlation Analysis (MFDXA). MFDXA is a meticulous application for computation of cross-correlation between simultaneously measured nonstationary time series. Significant difference in degree of multifractality is observed for wind energy, wind speed and wind direction. Wind direction is found to possess the highest degree of multifractality implying that the degree of complexity of wind direction is higher than wind speed or energy. Further strong cross-correlation between wind energy and wind direction is an indication that the direction of wind is one of the crucial factors in generation of wind energy. Thus, the cross-correlation analysis between wind energy - wind speed, and between wind energy - wind direction gives significant information about the scaling behavior, which may have necessary inputs towards optimization of wind power generation.

## 1. Introduction:

Wind energy, the kinetic energy form of the atmospheric air, when drives the propellers of the wind turbines produces one of the most environment friendly form of electrical energy. Wind power generation technology not only saves the environment from huge amount of carbon emissions and wastage of gallons of water a year, but also ensures the cheapest renewable energy at hand today (Medina et al., 2015). Owing to rapid population and economic growth, the demand for sustainable energy has increased manifold. Energy is also one of the crucial indicators of economic development. Wind energy along with the other forms of renewable energies, such as solar, tidal, geothermal etc. plays a major role as a low emitting power originator since last few decades (Calif et al., 2016). However, owing to turbulent behavior of wind, the produced wind power faces intermittent issues (Medina et al., 2015; Calif et al., 2013). Such unpredictable behavior of wind energy over short time scales is the main constraint when it is integrated with the electric grid (Calif et al., 2016; Shi et al., 2014; Ren et al., 2017).

With the rise in demand for clean and sustainable energy, research for developing clean energy technologies (Desideri and Yan 2012), and more innovating sustainable future energy systems (Yan et al., 2013) has stimulated in the recent times. Wind energy because of its competitive production cost and the ability to effectively overcome the environmental problems linked with the use of traditional energy sources, has become one of the most favorable sources of energy. As wind speed strongly influences the wind energy, exploring the structure of wind speed time series is important not only for better designing of wind power plants, but also to understand the dynamical mechanisms governing its variability more accurately (Telesca et al., 2016).

Wind is defined by its speed and direction which are non-independent parameters. One of the main components of air circulation is wind speed; thus, a low or zero value of wind speed fails to accurately measure the wind direction. Strong winds allow dispersion of atmospheric pollutants in the Atmospheric Boundary Layer, while calm winds promote its stagnation (Holzworth, 1972; Sharan et al., 1996; Malek et al., 2006), thus, it is important to understand wind speed fluctuation behavior (Plocoste and Pavon-Dominguez 2020).

Research on the intermittent character of wind power applying various mathematical models (Ngoc et al., 2011; Rahimi et al., 2013; Huang et al., 2014), aiming satisfactory solutions for the variability in short time scales of wind-speed with certain predictive approaches have been reported by several researchers (Dittner and Vasel, 2019; Melzer et al., 2019; Kim and Hur, 2020; Deng et al., 2020). As an integral part of wind energy forecasting technique, several attempts have been made to characterize the scaling properties of both wind speed and wind energy time series obtained from a number of wind-farms, situated worldwide with different generation capacities (Medina et al., 2015; Calif and Smith, 2014; Schmitt and Huang, 2016; González-Aparicio et al., 2017; Braun et al., 2020).

Several studies about wind speed time series suggest its nonlinear and nonstationary behavior (Govindan and Kantz, 2004; Kavasseri and Nagarajan, 2005; Calif and Schmitt, 2014). Their scaling features have been widely studied using the concept of fractals. The application of fractal methodology is based on the segmentation of a time series into self-similar parts and the exploration of the power-law behavior that reflects the scaling characteristics of the system examined (Tzanis et al., 2020). Chang et al., (2012) used box counting method to analyze wind speed time series at three wind farms in Taiwan under different climatic conditions and found an inverse correlation between wind speed and fractal dimension. de Oliveira Santos et al. (2012) studied the average and maximum hourly wind speed time series at four weather stations in Brazil using detrended fluctuation analysis (DFA), and found that both these two observables are characterized by almost identical power-law behavior, with two different scaling regimes and two different DFA scaling exponents. All these studies reported, modeled wind speed time series as monofractal, implying that a single scaling exponent (the fractal dimension D, the DFA scaling exponent) is sufficient to gather all the information about the dynamics of fluctuations of wind speed (Telesca et al., 2016). Monofractals can characterize homogeneous series, but for more complex and heterogeneous series different scaling exponents are required to describe the dynamics of the series which are composed of many interwoven fractal subsets into which the original series can be decomposed (Ivanov et al., 1999). These time series are called multifractals, and are characterized by spiky dynamics, with sudden and intense bursts of high frequency fluctuations (Meneveau and Sreenivasan 1991).

Multifractal analysis techniques have achieved the highest precision in the nonlinear scaling analysis through its applications in several disciplines such as, turbulence (Schaffner and Brown, 2015), finance (Ghosh et al., 2012a; Dutta et al., 2014), physiology (Ghosh et al., 2014; Ghosh et al., 2017a,b; Ghosh et al., 2018; Ghosh et al., 2019a), rainfall (Venugopal et al., 2006), geophysics (Kantelhardt et al., 2006; Ghosh et al., 2012b), linguistics (Ghosh et al., 2019b) etc. The applications of multifractal analysis techniques have even been extended to various domains of research in engineering and technology too (Ghosh et al., 2015; Ghosh et al., 2016; Zaghloul et al., 2018; Cekli et al., 2018; Yao et al., 2018; Raghunathan et al., 2020). Multifractality of wind speed time series were reported by the pioneers Govindan and Kantz (2004) and Kavasseri and Nagarajan (2005) in which the multifractal structure in long term correlations of these time series was described. Feng et al., (2009) observed different degrees of multifractality of daily wind speeds in China. Fortuna et al., (2014) applied multifractal detrended fluctuation analysis (MFDFA) to hourly wind speed time series at different weather stations in Italy and USA and found similar values of multiufractal width for the time series. de Figueirêdo et al., (2014) studied the mean and maximum of four wind speed time series in Brazil using MFDFA and reported persistent correlation of both the series with a larger multifractality for the maximum than for the mean. Piacquadio and de la Barra (2014) reported use of some important multifractal parameters of wind speed as local indicator of climate change.

Some other works on scaling analysis of wind energy generations along with wind speed, direction, pressure, kinetic energy, viscosity etc. (de Figueirêdo et al., 2014; Wang et al., 2016; Weerasinghe et al., 2016; Ali et al., 2016; Telesca et al., 2016; Kryuchkova et al., 2017; Strijhak et al., 2017; Laib et al., 2018; Cadenas et al., 2019; Ali et al., 2019) in light of multifractal analysis techniques have also been reported. Many researchers have also analyzed multifractal behavior of wind speed variations with various combinations of wind turbines and their applications in different arrays at different geographical locations (Ali et al., 2016; Telesca et al., 2016; Kryuchkova et al., 2017; Strijhak et al., 2017; Ali et al., 2019; Balkissoon et al., 2020). In a recent article Braun et al., (2020) demonstrated autocorrelation of wind power using DFA in an offshore wind farm in Germany. In 2021 Elsisi et al., (2021) proposed a neuro-fuzzy inference system (ANFIS) as an effective control technique for blade pitch control of the wind energy conservation system (WECS) instead of the conventional controllers. The work also suggested an effective strategy to prepare a sufficient dataset for training and testing of the ANFIS controller. Earlier in 2016, Promdee and Photong (2016) proposed certain significant modifications in the wind turbine designs, after conducting certain practical analysis on wind power, wind speed and wind direction angles. They investigated the generated voltage profile by adjusting the turbine blade angle at different wind speeds and wind angles. However, as far as the correlation analyses between the three major quantities i.e., wind energy, wind speed and wind direction (angle) are concerned, no studies have been reported so far to the best of our knowledge.

In order to see how variation in wind speed and direction affects the wind energy, the present work attempted to explore the correlation between continuous fluctuations in wind speed (m/s), and wind direction angle (º) with the generated wind energy (kWh) obtained from a "showcase" wind farm, using Multifractal Detrended Cross-Correlation Analysis (MFDXA). A "showcase" wind farm is a single wind farm with different wind turbine technologies. The main concern of such a farm is not only to ensure service facilities for the renewable energy sector, ensuring significant energy savings and higher efficiency, but also to facilitate R&D projects, training activities etc. Our study is mainly focused on the scaling behavior of ten years' daily wind data, from January 2010 to December 2019.

MFDXA proposed by Zhou (2008) is an advanced technique to explore multifractality among simultaneously recorded time series. The method has the advantage of exploring long-range cross-correlations between two simultaneously recorded nonstationary time series from a multifractal point of view. Another advantage of this method is that it uses a cross-correlation coefficient ($\gamma_x$) which gives the degree of correlation between two signals. For uncorrelated data, ($\gamma_x$) has a value 1; the lower the value of ($\gamma_x$) the more correlated is the data. Negative value of ($\gamma_x$) signifies very high degree of correlation between the signals, i.e., a large increment in one would more likely to follow a large increment of the other. MFDXA is a generalization of Detrended Cross-correlation Analysis (DCCA) that was proposed by Podobnik and Stanley (2008) to investigate power-law cross-correlations between two non-stationary time series. Application of MFDXA with significant high accuracy has been observed in analysing one- and two-dimensional binomial measures, multifractal random walks (MRWs) or financial prices and also in various fields of Science, Engineering, Medicine, Climate, Geography, Economics etc.

Tzanis et al., (2020) applied MFDXA to study the multifractal characteristics of global methane concentrations and remotely-sensed temperature anomalies of the global lower and mid-troposphere. Plocoste and Pavón-Domínguez (2020) studied the scaling behavior of wind speed and solar radiation using MFDXA. In 2015 dos Anjos et al., (2015) using DCCA showed the correlation between wind speed and solar radiation. MFDXA has also been used to explore the correlations between soil radon concentrations and land surface temperature (Kar et al., 2019), to study the cross-correlations between meteorological parameters and air pollution (Zhang et al., 2015), and also to examine the cross-correlations in sunspot numbers and river flow fluctuations (Hajian and Movahed 2010).

## 2. Materials and Methods:

The time series data for the present work was taken from the database of "Sotavento Galicia", S.A., Spain, a "showcase" wind farm, operating with 24 wind turbines, run by 5 different technologies and a total of 9 different machine models. The farm provides the raw wind data in 10 min resolutions and also average daily wind data.

### 2.1 Materials:

The experimental data for the study comprises of time series of average daily wind-energy (kWh), wind-speed (m/s) and wind direction (angle in degree) of ten years from January 2010 to December 2019 from database of 'Sotavento Galicia', S.A., Spain.

### 2.2 Methods:

The work reports analysis of two separate cross-correlation – 1) daily wind-energy (kWh) and corresponding wind-speed (m/s); and 2) daily wind-energy (kWh) and corresponding wind direction (angle of attack, in degree) applying MFDXA, a prescription by Zhou in 2008 (Zhou, 2008). MFDXA method is basically the integration of multifractal approach of Detrended Cross-Correlation Analysis (DCCA or DXA), developed by Podobnik and Stanley (2008).

MF-DXA is computed in the following steps:

- $x(i)$ and $y(i)$ are considered to be two non-stationary time series of length N, where $i = 1, \ldots \ldots \ldots, N$. The means of the above two series are given by

$$x_{avg} = \frac{1}{N}\sum_{i=1}^{N} x(i) \quad \text{and} \quad y_{avg} = \frac{1}{N}\sum_{i=1}^{N} y(i) \qquad (1)$$

  Also, the local trends of $x(i)$ and $y(i)$ are $x_{avg}$ and $y_{avg}$ respectively.

- The profiles of the underlying data series $x(i)$ and $y(i)$ are computed as

$$X(i) \equiv \sum_{k=1}^{i}[x(k) - x_{avg}] \quad \text{and}$$

$$Y(i) \equiv \sum_{k=1}^{i}[y(k) - y_{avg}] \quad \text{where } i = 1, \ldots \ldots N \qquad (2)$$

- The data series are then integrated to reduce noise level in the experimental records.
- Each of the integrated time series is divided into $N_s$ non-overlapping bins, where $N_s = \text{int}\left(\frac{N}{s}\right)$ and s be the length of the bin.
- Now since N may not necessarily be a multiple of s, a small part of the series remains left at the end. Therefore, to include that remaining part, the entire process is repeated again from the opposite end, thus $(N_s + N_s) = 2N_s$ bins are obtained altogether.
- The profiles within the ν-th bin are determined to be

$$x_\nu(i) = \sum_{k=1}^{i} x(l_\nu + k) \quad \text{and} \quad y_\nu(i) = \sum_{k=1}^{i} y(l_\nu + k), \text{ where } i = 1, \ldots, s; \text{ and } l_\nu = (\nu - 1)s$$

- For each bin, the least square linear fit is calculated and thereby the fluctuation function is obtained. The detrended covariance of each bin is calculated as

$$F(s,v) = \frac{1}{s}\sum_{k=1}^{s}\{Y[(v-1)s+k] - y_v(k)\} \times \{X[(v-1)s+k] - x_v(k)\}$$

  for each bin $v$, $v = 1, \ldots\ldots\ldots, N_s$ and

$$F(s,v) = \frac{1}{s}\sum_{i=1}^{s}\{Y[N-(v-N_s)s+i] - y_v(i)\} \times \{X[N-(v-N_s)s+i] - x_v(i)\}$$

  for $v = N_s + 1, \ldots\ldots\ldots, 2N_s$, where $x_v(i)$ and $y_v(i)$ are the least square fitted value in the bin $v$.

- The $q^{th}$ order detrended covariance $F_q(s)$ is obtained after averaging over $2N_s$ bins.

$$F_q(s) = \left\{\frac{1}{2N_s}\sum_{v=1}^{2N_s}[F(s,v)]^{q/2}\right\}^{1/q} \qquad (3)$$

  where q is an index which can take all possible values except zero, because in that case the factor 1/q blows up.

- The procedure can be repeated by varying the value of s. $F_q(s)$ increases with increase in value of s. If the series is long range power correlated, then $F_q(s)$ will show power law behavior

$$F_q(s) \propto s^{\lambda(q)}$$

- For such a scaling $\ln|F_q(s)|$ is proportional to $\ln|s|$, $\lambda(q)$ being the slope. Scaling exponent $\lambda(q)$ demonstrates the degree of the cross-correlation between the time series $x(i)$ and $y(i)$. The exponent $\lambda(q)$ depends on q. Logarithmic averaging process is adapted to find $\lambda(0)$

$$F_0(s) \equiv \exp\left\{\frac{1}{4N_s}\sum_{v=1}^{2N_s}\ln[F(s,v)]\right\} \sim s^{\lambda(0)} \qquad (4)$$

For q = 2, the method reduces to standard Detrended Cros-correlation Analysis (DCCA or DXA). Cross-correlations are monofractal when $\lambda(q)$ is independent of q and multifractal when $\lambda(q)$ is dependent on q.

Further, for positive q, $\lambda(q)$ describes the scaling behavior of the segments with large fluctuations; and for negative q, $\lambda(q)$ describes the scaling behavior of the segments with small fluctuations.

Different values of λ(q) has different meaning; where λ(q) = 0.5 means no cross-correlation, λ(q) > 0.5 depicts persistent long-range cross-correlations and λ(q) < 0.5 means anti-persistent cross-correlations (Yogev et al., 2005; Movahed and Hermanis, 2008).

Zhou found the subsequent relationship between two time series formulated from binomial measure of p-model (Zhou, 2008):

$$\lambda(q = 2) \approx [h_x(q = 2) + h_y(q = 2)]/2 \quad (5)$$

For monofractal Autoregressive Fractional Moving Average (ARFIMA) signals and EEG time series Podobnik and Stanley (2008) studied the behaviour of the above relationship putting q = 2.

Zhou too has revealed that the relationship is well founded for Multifractal Random Walks (MRW) and binomial measures generated from the p model (Mars et al., 1983, Hajian and Movahed, 2010). Though there are cases like daily price changes for DJIA and NASDAQ indices (Mars et al., 1983, Hajian and Movahed, 2010) where the relation cease to prevail, but it is still correct for q = 2 (Podobnik and Stanley, 2008; Shadkhoo and Jafari, 2009). The other example is that of time series created by using two uncoupled ARFIMA processes, each of both is auto-correlated, but there is no power-law cross correlation with a specific exponent (Podobnik and Stanley, 2008; Shadkhoo and Jafari, 2009).

- If auto-correlation function is:

$$C(\tau) = \langle [x(i+\tau) - \langle x \rangle][x(i) - \langle x \rangle] \rangle \sim \tau^{-\gamma} \quad (6)$$

then cross-correlation function can be written as (Hajian and Movahed, 2010)

$$C_x(\tau) = \langle [x(i+\tau) - \langle x \rangle][y(i) - \langle y \rangle] \rangle \sim \tau^{-\gamma_x} \quad (7)$$

where γ and $\gamma_x$ represents auto-correlation and cross-correlation exponents, respectively. Auto-correlation exponent is calculated from the DFA method where γ = 2 − 2h (q=2) (Kantelhardt et al., 2001; Movahed and Hermanis 2008).

The relationship between cross-correlation exponent, $\gamma_x$ and scaling exponent λ(q) can be calculated from Eq. (4) where $\gamma_x$ = 2 − 2λ(q=2). $\gamma_x$ = 1 signifies uncorrelated series. A lesser value of γ and $\gamma_x$ is manifestation of strong correlation is the series (Podobnik and Stanley, 2008). In general, λ (q) depends on q which is manifestation of multifractality of the cross-correlated series. For such dependence we can say that the two series are cross-correlated in various times scale. Another way to determine the multifractality is by calculating the classical multifractal scaling exponent τ(q). It is given by

$$\tau(q) = q\lambda(q) - 1 \quad (8)$$

To clarify the correlations further (between the two-time series as described above) the singularity spectrum, f(α), is calculated. This spectrum provides meaningful details regarding the spread of the degree of cross-correlation in various time scales. The method to characterize multifractality of cross-correlation between two series is to relate via a λ(q) Legendre transform, like in a single series (Feder, 1988; Peitgen et al., 2004).

$$\alpha = \lambda(q) + q\lambda'(q) \quad (9)$$

$$f(\alpha) = q[\alpha - \lambda(q)] + 1 \quad (10)$$

Here α depicts singularity strength or Hölder's exponent, and f(α) is dimension of the subset of the series which is specified by α. Hölder's exponent symbolizes monofractality. A multifractal series gives rise to a spectrum which is characterized by different values of α at different parts of the spectrum. To compute width of the spectrum, the fitted curve is extrapolated to zero as

Width w is computed as

$$w = \alpha_1 - \alpha_2 \quad (11)$$

with $f(\alpha_1)$ - $f(\alpha_2)$ = 0. The growth of the width of f(α) is a manifestation of complexity in the degree of multifractality of two coupled signals.

## 3. Results and Discussion:

The cross-correlation between the non-stationary time series: 1) daily wind energy (kWh) and corresponding wind speed (m/s); and 2) daily wind energy (kWh) and corresponding wind direction (angle in degree) fluctuation from Janaury 2010 – December 2019 has been studied using MFDXA methodology as described above.

All the signals are first transformed to obtain the integrated time series. These integrated time series are then divided into $N_s$ segments, where $N_s$ = int(N/s), N being the length of the time series. For q ranging between -10 to +10, the fluctuation function $F_q(s)$ is determined from Eqs. (3) and (4). q is varied in steps of 1. The plot of ln│$F_q(s)$│ vs. ln│s│ both for auto-correlated signals (i.e., wind energy, wind speed and wind direction) using MF-DFA and cross-correlated signals (i.e., wind energy – wind speed and wind energy – wind direction) using MF-DXA is portrayed in Fig.1. From the plot it can be observed that ln│$F_q(s)$│ increases with increasing ln│s│ thus relating $F_q(s)$ to s in a power law fashion. Dependence of ln│$F_q(s)$│ on ln│s│ indicates power law auto and cross-correlation between wind energy – wind speed and wind energy – wind direction. The slope of linear fit to ln│$F_q(s)$│ vs ln│s│ plots (Fig. 1) gives the values of the cross-correlation scaling exponent λ(q).

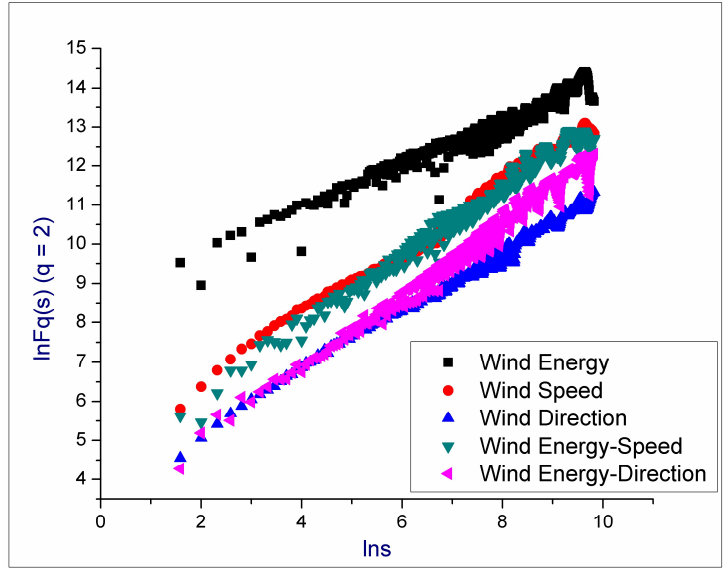

**Fig. 1:** Plot of ln│$F_q(s)$│ vs. ln│s│ at q = 2 for individual and cross-correlated datasets respectively (color online)

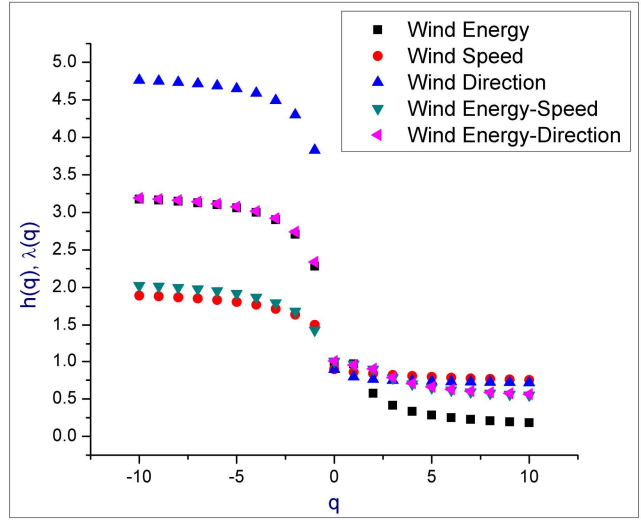

**Fig. 2:** Plot of h(q), λ(q) vs q for individual and cross-correlated datasets respectively (color online)

The relationship between the cross-correlation scaling exponent λ(q) and q in case of wind energy – wind speed and wind energy – wind direction is depicted in Fig. 2. In the same figure the variation of generalized Hurst exponent h(q) for different values of q obtained from MF-DFA is also shown for comparison. Both h(q) and λ(q) exhibits dependence on q which establishes the multifractality of wind energy, wind speed and wind direction and also of the cross-correlated series respectively, as for a monofractal series both h(q) and λ(q) are independent of q. The corresponding values of h(q=2) and λ(q=2), are listed in Table 1. From the values we can see that at q = 2, the cross-correlation scaling exponent λ(q) for wind energy – wind speed & wind energy – wind direction is greater than 0.5, which is an implication that long range cross-correlation and persistent properties exist in both the sets. Similarly for wind energy, wind speed and wind direction, the generalized Hurst exponent h (q =2) is noted to be greater than 0.5, which also implies the existence of long range auto-correlation and persistent properties in all the three sets. Liab et al., (2015) used MF-DFA to study the daily means of wind speed measured by 119 weather stations in different regions of Switzerland and reported persistent behaviour of almost all recorded wind speed time series. Zeng et al., (2016) applied multiscale multifractal analysis to estimate the scaling properties of a group of wind speed time series and found the time series to demonstrate strong multifractal properties and attributed the multifractality primarily due to long range correlations. Persistent behaviour of wind speed was also reported by Plocoste and Pavon-Dominguez (2020).

**Table 1:** Values of generalized Hurst exponent h(q) for individual set and scaling exponent λ(q) for cross-correlated set at q = 2.

| Energy | Speed | Direction | Energy-Speed | Energy-Direction |
| --- | --- | --- | --- | --- |
| $h_E$ (q=2) | $h_S$ (q=2) | $h_D$ (q=2) | $\lambda_{ES}$ (q=2) | $\lambda_{ED}$ (q=2) |
| 0.574 ± 0.004 | 0.838 ± 0.002 | 0.763 ± 0.002 | 0.898 ± 0.003 | 0.904 ± 0.003 |

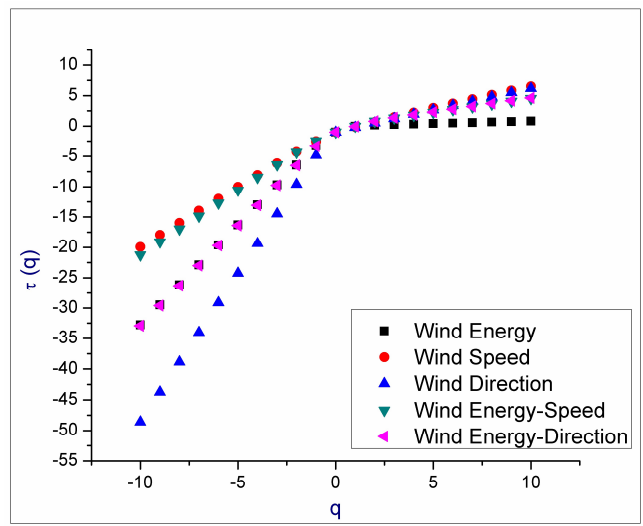

**Fig. 3:** Plot of τ(q) vs q for individual and cross-correlated datasets respectively (color online)

Fig. 3 represents variation of classical scaling exponent τ(q) against q. Nonlinearly dependence of τ(q) on q is observed for all the three sets viz. wind energy, wind speed and wind direction and also for the cross-correlated series of wind energy – wind speed & wind energy – wind direction. This nonlinear dependence is manifestation of multifractal nature of each of the individual and cross-correlated series. In case the dependence was linear the series would have been monofractal in nature.

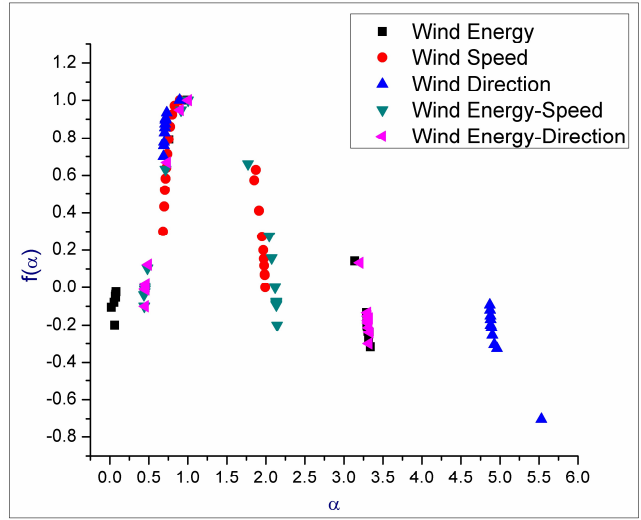

**Fig. 4:** Plot of f(α) vs α for individual and cross-correlated datasets respectively (color online)

Fig. 4 represents the plot of multifractal spectrum f(α) vs. Holder exponent α for all the individual and cross-correlated sets. The figure further confirms the multifractality of wind energy, wind speed, wind direction, and of the cross-correlated series of wind energy – wind speed and wind energy – wind direction. The corresponding multifractal width (W and $W_x$) of the multifractal spectrum is calculated by using equation 11. The value of multifractal width (W and Wx) for each of the individual and cross-correlated series is presented in Table 2. Significant variation in multifractal width (W) is observed, where wind direction has highest degree of multifractality compared to wind energy and wind speed. Since width is the measure of degree of multifractality or complexity, it can be inferred that wind direction has highest degree of complexity compared to wind speed and wind energy. Multifractal width ($W_x$) of the cross-correlated series wind energy – wind direction is greater than wind energy – wind speed. Thus, it is evident that the direction of the wind is crucial in determining the generation of wind power. Weerasinghe et al., (2016) in their study also observed high degree of multifractality for wind direction in contrast to wind speed for each season.

**Table 2:** Values of multifractal width W for individual data and $W_x$ for cross-correlated series

| Energy | Speed | Direction | Energy-Speed | Energy-Direction |
| --- | --- | --- | --- | --- |
| $W_E$ | $W_S$ | $W_D$ | $W_{x(ES)}$ | $W_{x(ED)}$ |
| 3.070 ± 0.046 | 1.421 ± 0.056 | 5.595 ± 0.512 | 1.654 ± 0.035 | 2.778 ± 0.037 |

In Table 3, the mean values of the auto-correlation exponent γ for the sets wind energy, wind speed & wind direction, and the corresponding cross-correlation exponent $γ_x$ for the sets wind

energy – wind speed & wind energy – wind direction are presented. As already stated earlier it is now known that a lower value of γ or $γ_x$ indicates higher degree of correlation. Strong auto-correlation is observed in case of both wind speed and wind direction. The values of cross-correlation exponent $γ_x$ implies strong cross-correlation between wind energy - wind speed and wind energy – wind direction.

**Table 3:** Values of auto-correlation exponent γ for individual series and cross-correlation exponent $γ_x$ for cross-correlated series (q = 2)

| Energy | Speed | Direction | Energy-Speed | Energy-Direction |
| --- | --- | --- | --- | --- |
| $γ_E$ | $γ_S$ | $γ_D$ | $γ_x$ (ES) | $γ_x$ (ED) |
| 0.852 ± 0.008 | 0.324 ± 0.005 | 0.474 ± 0.004 | 0.205 ± 0.006 | 0.193 ± 0.005 |

This strong cross–correlation signifies that if the wind speed changes then the wind energy will also change. Likewise, when wind direction changes the wind energy also changes accordingly, i.e if there is a change in one variable there is significant influence in the value of the other. Thus, from the values of cross-correlation exponent ($a_x$) it can be inferred that wind energy – wind direction is more cross-correlated than wind energy - wind speed. Thus, it is evident that wind direction has a significant influence in generation of wind energy which is also evident from the high value of multifractal width of wind direction. The results reported in the present work are consistent with the findings of Plocoste and Pavon-Dominguez (2020). They examined multifractal cross-correlation properties of wind speed and solar radiation in Guadeloupe archipelago. Using MF-DFA multifractality and persistent behaviour of wind speed and solar radiation was noted across all stations. Using MFDXA, cross-correlation analysis was performed among wind speed and solar radiation. Identical multifractal cross-correlation degree was observed for each site.

### 3.1. Discussion of Figures:

Fig 1 represents plot of the fluctuation function $F_q(s)$ against time scale s. The dependence of $\ln|F_q(s)|$ on $\ln|s|$ is manifestation of power law behavior i.e long range auto and cross-correlation exists in the time series. The values of the scaling exponent $λ(q)$ obtained from the linear fit of the plot of $\ln|F_q(s)|$ - $\ln|s|$ is presented in Fig. 2 which is manifestation of multifractality of the cross-corelated time series of wind energy – wind speed and wind energy – wind direction. The variation of Hurst exponent values, h(q) obtained from MF-DFA is shown in the same Fig. 2 for comparison between the auto and cross-correlated time series. The figure clearly presents the variation of both $λ(q)$ and h(q) with q i.e., the series are multifractal, as for a monofractal series both $λ(q)$ and h(q) are independent of q. For positive values of q, $λ(q)$ describes the scaling behavior of the segments with large fluctuations and for negative q, $λ(q)$ describes the scaling behavior of the segments with small fluctuations.

The variation of classing scaling exponent τ(q) with q as depicted in Fig. 3 is another way of characterizing the multifractality of the individual (wind energy, wind speed, wind direction) and cross-correlated (wind energy – wind speed & wind energy – wind direction) time series. A linear dependence of τ(q) on q would have meant that the series are monofractal.

The multifractal spectrum (Fig. 4) further characterizes multifractality of both individual and cross-correlated series. The degree of multifractality also denotes the degree of complexity of the series. The figure very clearly demonstrates that the degree of multifractality is the highest in case of wind direction thus implying its high degree of complexity. In other words, it can be said that the wind direction is one of the important factors which determines the optimal generation of wind power.

## 4. Conclusion:

The present analysis of the different wind time series for a period of ten years using MFDXA reveals the following important information:

a) For all sets at q = 2, both the scaling exponents λ(q) and h(q) are greater than 0.5 which confirms presence of both long-range auto and cross-correlation and persistent properties of the time series.

b) The values of multifractal width (W and $W_x$) give information about the existence of different degree of multifractality in all the sets. The degree of multifractality is highest for wind direction (5.595 ± 0.512) followed by wind energy (3.070 ± 0.046) and wind speed (1.421 ± 0.056), thus implying that wind direction has the highest degree of complexity. Thus, we can conclude that wind direction is one of the major factors which lead to optimal generation of power.

c) From the values of the correlation coefficients (γ and $γ_x$), the degrees of correlations can be quantified. Strong cross-correlations have been observed for both wind energy – wind direction ($γ_x$ = 0.193 ± 0.005) and wind energy – wind speed ($γ_x$ = 0.205 ± 0.006), where a lower value of $γ_x$ indicates a strong degree of cross-correlation. Strong cross-correlation implies that the change in wind direction has a significant influence in both the wind energy and wind speed. Higher degree of correlation of wind energy – wind direction compared to wind energy – wind speed also implies that maximum wind speed may not always lead to optimal wind energy; rather the direction is equally or more important for generating optimal wind energy.

d) The values of auto-correlation (γ) further suggest that wind speed (γ = 0.324 ± 0.005) is more auto-correlated than wind direction (0.474 ± 0.004) or wind energy (0.852 ± 0.008). From the values of γ and $γ_x$ it can be inferred that the cross-correlations between wind energy – wind speed and wind energy – wind direction are stronger than their auto-correlations.

e) Comparing $\gamma_{ED}$ and $\gamma_{ES}$ we may further conclude that wind energy is even more correlated with wind direction as compared to wind speed, which clearly directs that unless there is a very accurate angle of attack, i.e., the angle at which wind force strikes turbine blades at any speed, there cannot be optimal power generation.

In view of the existence of strong correlations between wind energy–wind directions and also between wind energy–wind speed, it may be concluded that by adjusting the turbine blade angles according to the dynamics of angle of attack at certain speeds, optimum wind energy can be harnessed for a particular season. Further, the turbine shaft angle may also be adjusted for different seasons subject to very large changes in the dynamics of wind flow.

The information obtained from the present work can be used in future to study the correlations of wind power with wind speed and wind direction in different seasons. Seasonal study is important as both wind speed and direction are not uniform but varies with different seasons throughout the year.

**Acknowledgement:** We would like to acknowledge 'Sotavento Galicia' from where we have obtained the data for the study.

Meneveau, C., and Sreenivasan, K. R. (1991). The multifractal nature of turbulent energy dissipation. *Journal of Fluid Mechanics*, 224, 429–84.

Movahed, M.S., and Hermanis, E. (2008). Fractal analysis of river flow fluctuations. *Physica A*, 387, 915–932.

Ngoc, P.D.N., Pham, T.T.H, Seddik, B., and Roye, D. (2011). Optimal management of wind intermittency in constrained electrical network, wind farm - Impact in Power System and Alternatives to Improve the Integration, Gastón Orlando Suvire, 109–144 (IntechOpen: London, UK, 2011).

Peitgen, H.O., Jürgens, H., and Saupe, D. (2004). Chaos and Fractals, 2nd ed. (Springer-Verlag: New York, US, 2004), p.p. 1–864.

Piacquadio, M., and de la Barra., A. (2014). Multifractal analysis of wind velocity data. *Energy for Sustainable Development*, 22, 48–56. Wind Power Special Issue.

Plocoste, T., and Pavon-Dominguez, P. (2020). Multifractal detrended cross-correlation analysis of wind speed and solar radiation. *Chaos*, 30, 113109.

Podobnik, B., and Stanley, H.E. (2008). Detrended cross-correlation analysis: a new method for analyzing two nonstationary time series. *Physical Review Letters*, 100, 084102.

Podobnik, B., Grosse, I., Horvatić, D., Ilic, S., Ivanov, P.Ch., and Stanley, H.E. (2009). Quantifying cross-correlations using local and global detrending approaches. *European Physical Journal B*, 71, 243–250.

Promdee, C., and Photong, C. (2016). Effects of wind angles and wind speeds on voltage generation of Savonius wind turbine with double wind tunnels. Procedia Computer Science, 86, 401–404. 2016 International Electrical Engineering Congress, iEECON2016, 2-4 March 2016, Chiang Mai, Thailand.

Raghunathan, M., George, N., Unni, V., Midhun, P., Reeja, K., and Sujith, R. (2020). Multifractal analysis of flame dynamics during transition to thermoacoustic instability in a turbulent combustor. *Journal of Fluid* Mechanics, 888, A14.

Rahimi, E., Rabiee, A., Aghaei, J., Muttaqi, K., and Nezhad, A.E. (2013). On the management of wind power intermittency. *Renewable and Sustainable Energy Reviews,* 28, 643–653.

Ren, G., Liu, J., Wan, J., Guo, Y., and Yu, D. (2017). Overview of wind power intermittency: Impacts, measurements, and mitigation solutions. *Applied Energy,* 204, 47–65.

Schaffner, D.A., and Brown, M.R. (2015). Multifractal and monofractal scaling in a Laboratory magnetohydrodynamic turbulence experiment. *The Astrophysical Journal*, 811, 1–7.

Schmitt, F.G., and Huang, Y. (2016). Stochastic analysis of scaling time series: from turbulence theory to applications. 1st ed. 1–226 (Cambridge University Press: Cambridge, UK, 2016).